\documentclass[reprint, amsmath, amssymb, aps, prl, showpacs]{revtex4-1}
\usepackage{graphicx}
\usepackage{epstopdf}

\begin{document}

\title{Photovoltage Bleaching in Bulk Heterojunction Solar Cells through \\ Occupation of the Charge Transfer State}%

\author{H.M. Shah}%
\author{A.D. Mohite}
\altaffiliation[Currently at ]{Los Alamos National Laboratory}
\author{T. Bansal}
\author{B.W. Alphenaar}
\email{brucea@louisville.edu}
\affiliation{Dept.\ of Elec.\ \& Comp.\ Eng.\ , University of Louisville, Louisville, KY, 40292}

\date{\today}%

\begin{abstract}
We observe a  strong peak  in the capacitive photocurrent of a MDMO-PPV / PCBM bulk heterojunction solar cell for excitation below the absorbance threshold energy. Illumination at the peak energy blocks charge capture at other wavelengths, and causes the photovoltage to drop dramatically. These results suggest that the new peak is due to a charge transfer state, which provides a pathway for charge separation and photocurrent generation in the solar cell.  
\end{abstract}

\maketitle
Organic photovoltaic (OPV) cells are cheaper and more easily fabricated than silicon solar cells and other inorganic photovoltaics. However, they also suffer from relatively low efficiency and time dependent output degradation.  Solving these problems requires an improved understanding of the unique charge generation mechanism in OPVs. The highest OPV efficiencies have been observed in bulk heterojunction (BHJ) solar cells.  BHJs consist of a mixture of two different organic materials \textemdash a $\pi$ conjugated polymer (which acts as an electron donor) and a fullerene based polymer (which acts as an electron acceptor)\cite{yu1, yu2, shah}.  Under illumination, excitons (or bound electron-hole pairs) are formed\cite{nunzi}. For photocurrent to be produced, the excitons must dissociate across the donor/acceptor interface and the resulting free carriers diffuse to the contacts. Recent experimental\cite{goris, jjbs, hall, trivngs}  and theoretical\cite{aryan} work suggests that the excitons dissociate via an intermediate charge transfer state (or exciplex\cite{morteani}) consisting of an electron hole pair bound across the donor / acceptor interface.  There is still considerable debate however as to the nature of this state, and what its role is in the charge dissociation process.

Here, we describe characterization of BHJ solar cells using capacitive photocurrent spectroscopy, (CPS) a novel spectroscopy technique developed in our laboratory\cite{mohite1} that is particularly sensitive to the exciton dissociation process.  CPS has been successfully used to distinguish between excitonic and free carrier states in individual carbon nanotubes\cite{mohite2, mohite3} and is able to detect exciton dissociation for carriers not captured by the electrical contacts.  Using CPS we are able to identify a photo-absorption state lying at an energy below the main exciton peak in an MDMO-PPV / PCBM solar cell. This peak has low absorbance, but very high dissociation efficiency, and its energy correlates well with previous observations of the charge transfer state\cite{hall}.  Illumination at the peak energy results in a decrease in the photovoltage signal by more than 70\%, while no decrease is observed under lower or higher energy illumination.  This strongly suggests that this state has a significant role in the charge dissociation process.

 \begin{figure}
 \includegraphics{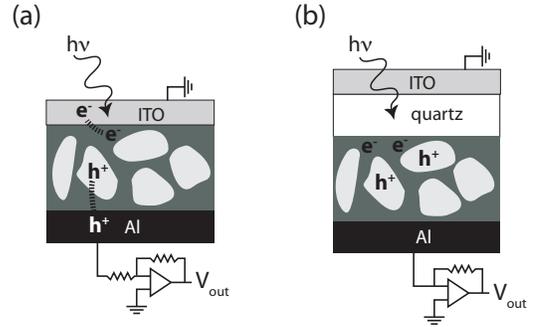}
\caption{(a) Experimental set-up for the photovoltage measurement. The active region consists of a mixture of an electron donor (light regions) and an electron acceptor (dark regions). Electron-hole pairs excited by the incident light dissociate, and diffuse to the contacts where they are detected as an in-phase voltage.  (b) Experimental set-up for the capacitive photocurrent measurement.  The quartz dielectric blocks charge capture by the contacts.  Dissociation of electron-hole pairs in the polymer results in an out-of phase voltage which is detected by a current amplifier which forms a virtual ground at the Al contact. }
 \end{figure}

A comparison of  standard photovoltage and capacitive photocurrent measurement techniques is shown in Figs. 1(a) and (b), respectively. In both cases, the active region consists of a 1:4 mixture of  MDMO-PPV:PCBM.  In the standard photovoltage measurement, electrical contacts are made to the top and bottom of the sample, using Al/LiF and ITO/PEDOT:PSS, respectively. Light incident on the polymer (through the transparent ITO contact) excites electron-hole pairs into excitonic states.  Some fraction of the excitons dissociate into free carriers, and some fraction of these then diffuse to the contacts to be detected as a photovoltage.  

 \begin{figure}
 \includegraphics{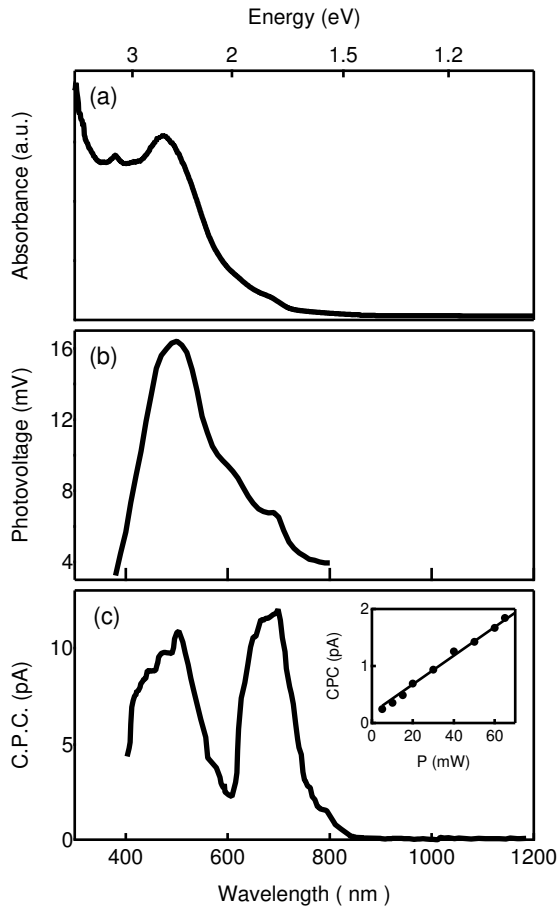}
\caption{Comparison of the (a) absorbance, (b) standard photovoltage and (c) capacitive photocurrent for a  MDMO-PPV:PCBM bulk heterojunction solar cell. The inset in (c) shows the excitation power dependence of the low energy feature in the capacitive photocurrent, measured for a second device (whose spectrum is  shown in Fig. 4(d)) at an excitation of 688 nm.}
 \end{figure}

In the capacitive photocurrent measurement, the ITO contact is separated from the polymer by an insulating quartz layer.  This blocks the flow of dc current to the contact;  however, the probe remains sensitive to charge that dissociates and separates (due to the built-in potential) to form a net dipole moment in the polymer layer. For a modulated light source this produces an out-of-phase ac voltage which can be measured with respect to the isolated ITO contact.  An advantage of this technique is that it is sensitive to charge that dissociates, but is unable to diffuse to the contacts.  By contrast, in the standard photovoltage measurement the out-of-phase signal is dominated by the in-phase (or dc) signal due to the diffusion of carriers to the contacts.  Carriers which dissociate, but do not make it to the contact remain undetected. 

Figure 2(a) shows the absorbance spectrum of a 100 nm thick layer of 1:4 MDMO-PPV:PCBM,  spin coated onto a glass slide.  The measurements were performed using a Perkin Elmer Lambda 950 UV-VIS spectrometer under atmospheric conditions.  A peak in the absorbance is observed at 2.6 eV (477 nm), and drops-off at lower energy.  This peak has been attributed to the ground excitonic state in the MDMO-PPV . A second smaller peak associated with the ground state exciton of the PCBM lies at 3.28 eV (378 nm)\cite{cook}. Fig. 2(b) shows the open circuit photovoltage of the BHJ measured using the two-contact set-up shown in Fig. 1(a) with the sample in vacuum.  The sample is illuminated by a tungsten halogen white light source (Newport Q-T-Halogen, 1 kW ) resolved by a monochromator (Acton Research, SpectraPro 500i) over the wavelength range of 380 - 800 nm. The incident light is chopped at low frequency (13 Hz) and the photovoltage detected with a  lock-in amplifier.  The photovoltage roughly correlates with the absorbance (showing a peak at 2.6 eV), and is similar to photocurrent and photovoltage measurements of the MDMO-PPV:PCBM system described in the literature\cite{hoppe, shah}.  

Figure 2(c) shows the results of the capacitive photocurrent measurement of the BHJ, using the set-up shown in Fig. 1(b).  The capacitor structure is made by spin coating a 100 nm film of 1:4 (MDMO: PPV-PCBM) onto a quartz slide with an ITO contact on the opposite side. An aluminum contact is then deposited onto the polymer film.  The polymer and ITO function as the positive and negative electrodes of the capacitor, with the quartz functioning as the capacitor dielectric. The sample is anchored to a copper block within an optical access flow cryostat and kept at vacuum. Illumination is done with an optical parametric amplifier (OPA) excited by a pulsed Ti-Sapphire regenerative amplifier. This produces tunable excitation between 0.4 and 2.4 eV. The pulse width is 120 fs with a repetition rate of 1 kHz, and the output power is kept constant at 14 mW.  The capacitive photocurrent is detected by passing the output into a current amplifier and then measuring the out-of-phase signal using a lock-in amplifier. 

The capacitive photocurrent spectrum shows a peak at 2.6 eV, similar to that observed in the absorbance and standard photovoltage.  However, a second peak is also observed at 1.77 eV ( 699 nm)  whose magnitude is even larger than the ground state exciton peak.    Comparison to Figs. 2(a) and 2(b) shows that evidence for the low energy feature can also be observed in the absorbance and standard photovoltage.  In both cases, a small deflection is observed near 1.77 eV.   However, the feature is much stronger in the capacitive photocurrent spectrum. Since evidence for the new feature is observed in all three measurements, its appearance is clearly not dependent on the use of a femtosecond pulsed laser (such as a multiple photon transition) or to an anomaly of the capacitive photocurrent technique. 
To further confirm that the low energy CPS peak is not due to two-photon absorption, the magnitude of the peak was measured as a function of laser power. As shown in the inset to Fig. 2(c), the magnitude of the peak increases approximately linearly with increasing power, indicating that it is due to a single photon absorption process. 

The observation of a sub-absorption threshold feature in the CPS signal follows reports from a number of groups who have observed sub-gap features in the photoexcitation spectrum of BHJ solar cells using a variety of techniques (including absorption\cite{goris}, photothermal deflection\cite{jjbs}, photoluminescence\cite{hall}, electroluminescence\cite{trivngs} and electroabsorption\cite{holt}).  Photoluminescence measurements of MDMO:PPV-PCBM show a broad sub-gap peak centered at 1.65 eV\cite{hall}, close in energy to the CPS peak that we observe.  A question that remains is to what extent the observed sub-gap states contribute to the charge dissociation process.  From Fig. 2(a), it is clear that the absorbance is relatively weak at the low energy CPS peak.  The large magnitude of the CPS signal must mean that charge carriers photoexcited at this energy have a very high dissociation efficiency.  However, most of the dissociated carriers do not diffuse far enough to be captured by the contacts, (as indicated by relatively weak signal in the standard photovoltage measurement). It is still possible, though, that the low energy state provides a pathway through which higher energy excitons can dissociate into free carriers.  If so, it is expected that the occupation of the low energy state would reduce its availability for the dissociation of higher energy excitons, and that this would then lead to a reduction in the photovoltage.  

To test this hypothesis, we performed a series of standard photovoltage measurements while exposing the sample to fixed wavelength light.   The photovoltage is measured as a function of wavelength using light from the monochrometer (probe beam).  For each measurement, the sample is also exposed to fixed wavelength light from the OPA laser (pump beam).  The results are plotted in Fig. 3 for four different pump beam wavelengths (solid lines) and for the pump beam blocked (dashed line). The magnitude of the standard photovoltage decreases under exposure to the OPA light, by an amount that is strongly dependent on the wavelength of the pump beam. The maximum change is observed for a pump wavelength of 699 nm (1.77 eV), while almost no change is observed for a pump wavelength of 474 nm (2.6 eV). 

Figure 4(a) plots the percent change in the photovoltage signal (integrated over the 380-800 nm probe beam wavelength range) as a function of the pump beam wavelength. As stated above, the greatest change in the photovoltage is centered at an excitation energy of 1.77 eV (699 nm) where a reduction of 71\% is observed. For pump beam energies on either side of this value, the bleaching is reduced, forming a pronounced minimum in the photovoltage signal.   A comparison to the capacitive photocurrent spectrum (reproduced from Fig 2(c) in Fig. 4(b)) shows that the wavelength dependence of the bleaching directly correlates with the longer wavelength feature in the CPS.   To prove that this result is not specific to a particular device, Figs. 4(c) and 4(d) show results of the same measurement performed on a second device structure. Although there is some variation in the detailed spectrum, the main result is reproduced  \textemdash a large, sub-absorbance threshold peak is observed in the CP spectrum, and a decrease in the photovoltage is observed when the sample is exposed to light at the peak energy.  It is interesting to note that in the second device, a pair of sub-threshold peaks are observed, however, only one is reproduced in the bleaching measurements.  This suggests that lowest energy peak is not involved in the charge dissociation process, and is perhaps due to a sample specific defect state. 

 \begin{figure}
 \includegraphics{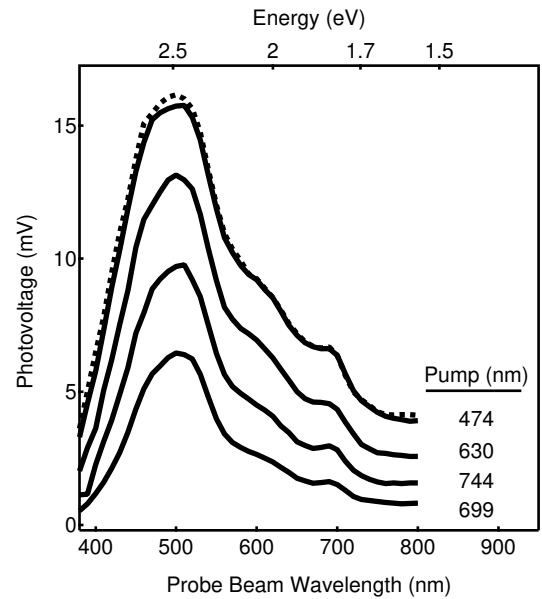}
\caption{Standard photovoltage spectra, measured as a function of probe beam wavelength while under illumination from a second pump beam.  Results are shown for four different pump beam wavelengths.  Also included are results for the pump beam blocked. (dashed line).}
 \end{figure}

It is striking that such large photovoltage bleaching occurs under exposure to long wavelength light even though the absorbance in this regime is relatively weak (see Fig. 2(a)).  In contrast, little or no bleaching is observed under exposure to short wavelength light corresponding to the main absorbance peak.   This demonstrates that the effect is not simply due to heating of the sample through absorption of the pump beam energy.  Instead, the bleaching must somehow be caused by the occupation of the 1.77 eV state. One possibility is that filling of the 1.77 eV state and the subsequent charge dissociation creates an electric field which blocks the flow of  additional charge to the contacts.  To test this, we measured the voltage directly by replacing the current amplifier used in the capacitive photocurrent measurement with a voltage amplifier.  We observe that the photocurrent peak corresponds to a voltage difference of only 30 $\mu$V, which is far less than the change of 10 mV observed in the standard photovoltage, making this explanation unlikely. In addition, no bleaching is observed under exposure to high energy light even though a similar or larger blocking voltage is created.

These results can be understood in terms of exciton dissociation through an interfacial charge transfer state, as has been described in the literature\cite{goris, jjbs, hall, trivngs, aryan,morteani}.  Here, the long wavelength peak corresponds to an intermediate state needed for the efficient dissociation of photo generated excitons, while the main absorbance peak (at 2.6 eV) corresponds to the ground exciton state of the MDMO-PPV.  The majority of absorbance occurs into the MDMO-PPV and PCBM exciton states. These have long diffusion lengths, but also long dissociation times\cite{muller}.  At the interface between the MDMO-PPV and PCBM, charge transfer excitons are formed with fast dissociation times\cite{sariciftci}.  However, these states also have short diffusion lengths and a small absorption cross section.  Charge pairs in the bulk exciton states diffuse to the interface, where dissociation occurs. If the interface is sufficiently close to the contact, a photovoltage is produced.  In our bleaching experiment, we populate the interfacial states, blocking the dissociation of the main excitonic states. This reduces the photovoltage by removing the pathway for efficient charge dissociation. 

\begin{figure}
 \includegraphics{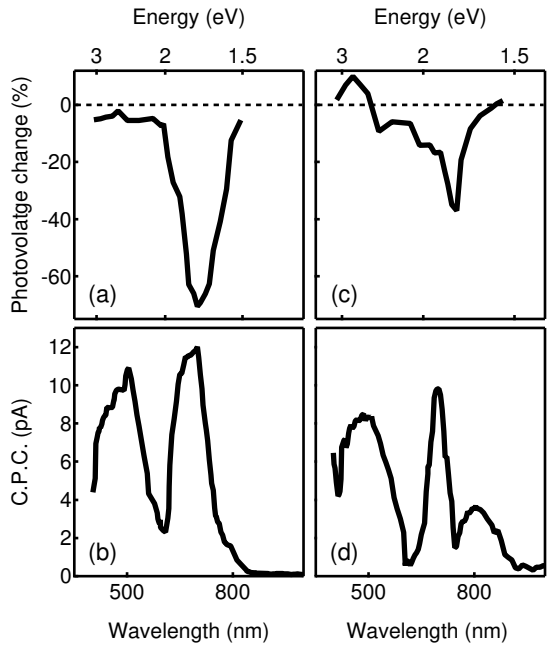}
\caption{(a) Percent change in the integrated photovoltage signal as a function of pump beam wavelength. Comparison to the capacitive photocurrent signal (reproduced in (b)) shows that the dip in the photovoltage occurs for the pump beam wavelength at which the sub-threshold peak is observed. (c) and (d) show similar results for a second MDMO-PPV:PCBM solar cell. }
 \end{figure}

 In conclusion, using the capacitive photocurrent technique we are able to resolve a low energy feature in the photo-excitation spectrum of BHJ solar cells that is much more weakly observed in standard absorbance and photovoltage spectroscopy.  The state has a large dissociation rate and a low absorbance cross-section compared to states at higher energy. Bleaching of the photovoltage signal is observed for illumination at the low energy feature suggesting that filling this state impedes the charge dissociation process. The experimental results counterintuitively demonstrate that increasing the amount of light on a BHJ solar cell can actually cause the output to go down. This implies that the solar cell efficiency could be improved by filtering out the light over a narrow band of wavelengths corresponding to the interfacial state energies.  

\bibliography{OPVtext2}

\providecommand{\noopsort}[1]{}\providecommand{\singleletter}[1]{#1}%
\begin{thebibliography}{10}%
\makeatletter
\providecommand \@ifxundefined [1]{%
 \ifx #1\undefined \expandafter \@firstoftwo
 \else \expandafter \@secondoftwo
\fi
}%
\providecommand \@ifnum [1]{%
 \ifnum #1\expandafter \@firstoftwo
 \else \expandafter \@secondoftwo
\fi
}%
\providecommand \enquote [1]{``#1''}%
\providecommand \bibnamefont  [1]{#1}%
\providecommand \bibfnamefont [1]{#1}%
\providecommand \citenamefont [1]{#1}%
\providecommand\href[0]{\@sanitize\@href}%
\providecommand\@href[1]{\endgroup\@@startlink{#1}\endgroup\@@href}%
\providecommand\@@href[1]{#1\@@endlink}%
\providecommand \@sanitize [0]{\begingroup\catcode`\&12\catcode`\#12\relax}%
\@ifxundefined \pdfoutput {\@firstoftwo}{%
 \@ifnum{\z@=\pdfoutput}{\@firstoftwo}{\@secondoftwo}%
}{%
 \providecommand\@@startlink[1]{\leavevmode\special{html:<a href="#1">}}%
 \providecommand\@@endlink[0]{\special{html:</a>}}%
}{%
 \providecommand\@@startlink[1]{%
  \leavevmode
  \pdfstartlink
   attr{/Border[0 0 1 ]/H/I/C[0 1 1]}%
   user{/Subtype/Link/A<</Type/Action/S/URI/URI(#1)>>}%
  \relax
 }%
 \providecommand\@@endlink[0]{\pdfendlink}%
}%
\providecommand \url  [0]{\begingroup\@sanitize \@url }%
\providecommand \@url [1]{\endgroup\@href {#1}{\urlprefix}}%
\providecommand \urlprefix [0]{URL }%
\providecommand \Eprint[0]{\href }%
\@ifxundefined \urlstyle {%
  \providecommand \doi [1]{doi:\discretionary{}{}{}#1}%
}{%
  \providecommand \doi [0]{doi:\discretionary{}{}{}\begingroup
  \urlstyle{rm}\Url }%
}%
\providecommand \doibase [0]{http://dx.doi.org/}%
\providecommand \Doi[1]{\href{\doibase#1}}%
\providecommand \bibAnnote [3]{%
  \BibitemShut{#1}%
  \begin{quotation}\noindent
    \textsc{Key:}\ #2\\\textsc{Annotation:}\ #3%
  \end{quotation}%
}%
\providecommand \bibAnnoteFile [2]{%
  \IfFileExists{#2}{\bibAnnote {#1} {#2} {\input{#2}}}{}%
}%
\providecommand \typeout [0]{\immediate \write \m@ne }%
\providecommand \selectlanguage [0]{\@gobble}%
\providecommand \bibinfo [0]{\@secondoftwo}%
\providecommand \bibfield [0]{\@secondoftwo}%
\providecommand \translation [1]{[#1]}%
\providecommand \BibitemOpen[0]{}%
\providecommand \bibitemStop [0]{}%
\providecommand \bibitemNoStop [0]{.\EOS\space}%
\providecommand \EOS [0]{\spacefactor3000\relax}%
\providecommand \BibitemShut [1]{\csname bibitem#1\endcsname}%
\bibitem{yu1}%
  \BibitemOpen
  \bibfield{author}{%
  \bibinfo {author} {\bibfnamefont{G.}~\bibnamefont{Yu}}, \bibinfo {author}
  {\bibfnamefont{J.}~\bibnamefont{Gao}}, \bibinfo {author}
  {\bibfnamefont{J.}~\bibnamefont{Hummelen}}, \bibinfo {author}
  {\bibfnamefont{F.}~\bibnamefont{Wudl}},\ and\ \bibinfo {author}
  {\bibnamefont{A.J.Heeger}},\ }%
  \bibfield{journal}{%
  \bibinfo {journal} {Science}\ }%
  \textbf{\bibinfo {volume} {270}},\ \bibinfo {pages} {1789} (\bibinfo {year}
  {1995})%
  \bibAnnoteFile{NoStop}{yu1}%
\bibitem{yu2}%
  \BibitemOpen
  \bibfield{author}{%
  \bibinfo {author} {\bibfnamefont{G.}~\bibnamefont{Yu}}\ and\ \bibinfo
  {author} {\bibfnamefont{A.}~\bibnamefont{Heeger}},\ }%
  \bibfield{journal}{%
  \bibinfo {journal} {J.\ Appl.\ Phys.}\ }%
  \textbf{\bibinfo {volume} {78}},\ \bibinfo {pages} {4510} (\bibinfo {year}
  {1995})%
  \bibAnnoteFile{NoStop}{yu2}%
\bibitem{shah}%
  \BibitemOpen
  \bibfield{author}{%
  \bibinfo {author} {\bibfnamefont{S.~E.}\ \bibnamefont{Shaheen}}, \bibinfo
  {author} {\bibfnamefont{C.~J.}\ \bibnamefont{Brabec}}, \bibinfo {author}
  {\bibfnamefont{N.~S.}\ \bibnamefont{Sariciftci}}, \bibinfo {author}
  {\bibfnamefont{F.}~\bibnamefont{Padinger}}, \bibinfo {author}
  {\bibfnamefont{T.}~\bibnamefont{Fromherz}},\ and\ \bibinfo {author}
  {\bibfnamefont{J.~C.}\ \bibnamefont{Hummelen}},\ }%
  \bibfield{journal}{%
  \bibinfo {journal} {Appl.\ Phys.\ Lett}\ }%
  \textbf{\bibinfo {volume} {78}},\ \bibinfo {pages} {841} (\bibinfo {year}
  {2001})%
  \bibAnnoteFile{NoStop}{shah}%
\bibitem{nunzi}%
  \BibitemOpen
  \bibfield{author}{%
  \bibinfo {author} {\bibfnamefont{J.}~\bibnamefont{Nunzi}},\ }%
  \bibfield{journal}{%
  \bibinfo {journal} {C. R. Physique}\ }%
  \textbf{\bibinfo {volume} {3}},\ \bibinfo {pages} {523} (\bibinfo {year}
  {2002})%
  \bibAnnoteFile{NoStop}{nunzi}%
\bibitem{goris}%
  \BibitemOpen
  \bibfield{author}{%
  \bibinfo {author} {\bibfnamefont{L.}~\bibnamefont{Goris}}, \bibinfo {author}
  {\bibfnamefont{A.}~\bibnamefont{Poruba}}, \bibinfo {author}
  {\bibfnamefont{L.}~\bibnamefont{Hod\'{a}kova}}, \bibinfo {author}
  {\bibfnamefont{M.}~\bibnamefont{Van\u{e}\u{c}ek}}, \bibinfo {author}
  {\bibfnamefont{K.}~\bibnamefont{Haenen}}, \bibinfo {author}
  {\bibfnamefont{M.}~\bibnamefont{Nesl\'{a}dek}}, \bibinfo {author}
  {\bibfnamefont{P.}~\bibnamefont{Wagner}}, \bibinfo {author}
  {\bibfnamefont{D.}~\bibnamefont{Vanderzande}}, \bibinfo {author}
  {\bibfnamefont{L.~D.}\ \bibnamefont{Schepper}},\ and\ \bibinfo {author}
  {\bibfnamefont{J.}~\bibnamefont{Manca}},\ }%
  \bibfield{journal}{%
  \bibinfo {journal} {Appl.\ Phys.\ Lett.}\ }%
  \textbf{\bibinfo {volume} {88}},\ \bibinfo {pages} {052113} (\bibinfo {year}
  {2006})%
  \bibAnnoteFile{NoStop}{goris}%
\bibitem{jjbs}%
  \BibitemOpen
  \bibfield{author}{%
  \bibinfo {author} {\bibfnamefont{J.~J.}\ \bibnamefont{Benson-Smith}},
  \bibinfo {author} {\bibfnamefont{L.}~\bibnamefont{Goris}}, \bibinfo {author}
  {\bibfnamefont{K.}~\bibnamefont{Vandewal}}, \bibinfo {author}
  {\bibfnamefont{K.}~\bibnamefont{Haenen}}, \bibinfo {author}
  {\bibfnamefont{J.~V.}\ \bibnamefont{Manca}}, \bibinfo {author}
  {\bibfnamefont{D.}~\bibnamefont{Vanderzande}}, \bibinfo {author}
  {\bibfnamefont{D.~D.~C.}\ \bibnamefont{Bradley}},\ and\ \bibinfo {author}
  {\bibfnamefont{J.}~\bibnamefont{Nelson}},\ }%
  \bibfield{journal}{%
  \bibinfo {journal} {Adv.\ Funct.\ Mater}\ }%
  \textbf{\bibinfo {volume} {17}},\ \bibinfo {pages} {451} (\bibinfo {year}
  {2007})%
  \bibAnnoteFile{NoStop}{jjbs}%
\bibitem{hall}%
  \BibitemOpen
  \bibfield{author}{%
  \bibinfo {author} {\bibfnamefont{M.}~\bibnamefont{Hallermann}}, \bibinfo
  {author} {\bibfnamefont{I.}~\bibnamefont{Kriegel}}, \bibinfo {author}
  {\bibfnamefont{E.~D.}\ \bibnamefont{Como}}, \bibinfo {author}
  {\bibfnamefont{J.}~\bibnamefont{Berger}}, \bibinfo {author}
  {\bibfnamefont{E.}~\bibnamefont{von Hauff}},\ and\ \bibinfo {author}
  {\bibfnamefont{J.}~\bibnamefont{Feldmann}},\ }%
  \bibfield{journal}{%
  \bibinfo {journal} {Adv.\ Funct.\ Mater.}\ }%
  \textbf{\bibinfo {volume} {19}},\ \bibinfo {pages} {3662} (\bibinfo {year}
  {2009})%
  \bibAnnoteFile{NoStop}{hall}%
\bibitem{trivngs}%
  \BibitemOpen
  \bibfield{author}{%
  \bibinfo {author} {\bibfnamefont{K.}~\bibnamefont{Tvingstedt}}, \bibinfo
  {author} {\bibfnamefont{K.}~\bibnamefont{Vandewal}}, \bibinfo {author}
  {\bibfnamefont{A.}~\bibnamefont{Gadisa}}, \bibinfo {author}
  {\bibfnamefont{F.}~\bibnamefont{Zhang}}, \bibinfo {author}
  {\bibfnamefont{J.}~\bibnamefont{Manca}},\ and\ \bibinfo {author}
  {\bibfnamefont{O.}~\bibnamefont{Ingan{\"{a}}s}},\ }%
  \bibfield{journal}{%
  \bibinfo {journal} {J. Am.\ Chem.\ Soc.}\ }%
  \textbf{\bibinfo {volume} {131}},\ \bibinfo {pages} {11820} (\bibinfo {year}
  {2009})%
  \bibAnnoteFile{NoStop}{trivngs}%
\bibitem{aryan}%
  \BibitemOpen
  \bibfield{author}{%
  \bibinfo {author} {\bibfnamefont{K.}~\bibnamefont{Aryanpour}}, \bibinfo
  {author} {\bibfnamefont{D.}~\bibnamefont{Psiachos}},\ and\ \bibinfo {author}
  {\bibfnamefont{S.}~\bibnamefont{Mazumdar}},\ }%
  \bibfield{journal}{%
  \bibinfo {journal} {Phys.\ Rev.\ B}\ }%
  \textbf{\bibinfo {volume} {81}},\ \bibinfo {pages} {085407} (\bibinfo {year}
  {2010})%
  \bibAnnoteFile{NoStop}{aryan}%
\bibitem{morteani}%
  \BibitemOpen
  \bibfield{author}{%
  \bibinfo {author} {\bibfnamefont{A.}~\bibnamefont{Morteani}}, \bibinfo
  {author} {\bibnamefont{P.Sreearunothai}}, \bibinfo {author}
  {\bibnamefont{L.M.Herz}}, \bibinfo {author}
  {\bibfnamefont{R.}~\bibnamefont{Friend}},\ and\ \bibinfo {author}
  {\bibfnamefont{C.}~\bibnamefont{Silva}},\ }%
  \bibfield{journal}{%
  \bibinfo {journal} {Phys.\ Rev.\ Lett}\ }%
  \textbf{\bibinfo {volume} {92}},\ \bibinfo {pages} {247402} (\bibinfo {year}
  {2004})%
  \bibAnnoteFile{NoStop}{morteani}%
\bibitem{mohite1}%
  \BibitemOpen
  \bibfield{author}{%
  \bibinfo {author} {\bibfnamefont{A.~D.}\ \bibnamefont{Mohite}}, \bibinfo
  {author} {\bibfnamefont{S.}~\bibnamefont{Chakraborty}}, \bibinfo {author}
  {\bibfnamefont{P.}~\bibnamefont{Gopinath}}, \bibinfo {author}
  {\bibfnamefont{G.~U.}\ \bibnamefont{Sumanasekera}},\ and\ \bibinfo {author}
  {\bibfnamefont{B.~W.}\ \bibnamefont{Alphenaar}},\ }%
  \bibfield{journal}{%
  \bibinfo {journal} {Appl.\ Phys.\ Lett.}\ }%
  \textbf{\bibinfo {volume} {86}},\ \bibinfo {pages} {061114} (\bibinfo {year}
  {2005})%
  \bibAnnoteFile{NoStop}{mohite1}%
\bibitem{mohite2}%
  \BibitemOpen
  \bibfield{author}{%
  \bibinfo {author} {\bibfnamefont{A.}~\bibnamefont{Mohite}}, \bibinfo {author}
  {\bibnamefont{T.S.Santos}}, \bibinfo {author} {\bibnamefont{J.S.Moodera}},\
  and\ \bibinfo {author} {\bibfnamefont{B.}~\bibnamefont{Alphenaar}},\ }%
  \bibfield{journal}{%
  \bibinfo {journal} {Nature Nanotechnology}\ }%
  \textbf{\bibinfo {volume} {4}},\ \bibinfo {pages} {425} (\bibinfo {year}
  {2009})%
  \bibAnnoteFile{NoStop}{mohite2}%
\bibitem{mohite3}%
  \BibitemOpen
  \bibfield{author}{%
  \bibinfo {author} {\bibnamefont{A.D.Mohite}}, \bibinfo {author}
  {\bibfnamefont{P.}~\bibnamefont{Gopinath}}, \bibinfo {author}
  {\bibfnamefont{H.}~\bibnamefont{Shah}},\ and\ \bibinfo {author}
  {\bibfnamefont{B.}~\bibnamefont{Alphenaar}},\ }%
  \bibfield{journal}{%
  \bibinfo {journal} {Nano Lett.}\ }%
  \textbf{\bibinfo {volume} {8}},\ \bibinfo {pages} {142} (\bibinfo {year}
  {2008})%
  \bibAnnoteFile{NoStop}{mohite3}%
\bibitem{cook}%
  \BibitemOpen
  \bibfield{author}{%
  \bibinfo {author} {\bibfnamefont{S.}~\bibnamefont{Cook}}, \bibinfo {author}
  {\bibfnamefont{H.}~\bibnamefont{Ohkita}}, \bibinfo {author}
  {\bibfnamefont{Y.}~\bibnamefont{Kim}}, \bibinfo {author}
  {\bibfnamefont{J.}~\bibnamefont{Benson-Smith}}, \bibinfo {author}
  {\bibnamefont{D.D.C.Bradley}},\ and\ \bibinfo {author}
  {\bibfnamefont{J.}~\bibnamefont{Durrant}},\ }%
  \bibfield{journal}{%
  \bibinfo {journal} {Chem.\ Phys.\ Lett.}\ }%
  \textbf{\bibinfo {volume} {445}},\ \bibinfo {pages} {276} (\bibinfo {year}
  {2007})%
  \bibAnnoteFile{NoStop}{cook}%
\bibitem{hoppe}%
  \BibitemOpen
  \bibfield{author}{%
  \bibinfo {author} {\bibfnamefont{H.}~\bibnamefont{Hoppe}}, \bibinfo {author}
  {\bibfnamefont{M.}~\bibnamefont{Niggemann}}, \bibinfo {author}
  {\bibfnamefont{C.}~\bibnamefont{Winder}}, \bibinfo {author}
  {\bibfnamefont{J.}~\bibnamefont{Kraut}}, \bibinfo {author}
  {\bibfnamefont{R.}~\bibnamefont{Hiesgen}}, \bibinfo {author}
  {\bibfnamefont{A.}~\bibnamefont{Hinsch}}, \bibinfo {author}
  {\bibfnamefont{D.}~\bibnamefont{Meissner}},\ and\ \bibinfo {author}
  {\bibfnamefont{N.}~\bibnamefont{Sariciftci}},\ }%
  \bibfield{journal}{%
  \bibinfo {journal} {Adv.\ Funct.\ Mater.}\ }%
  \textbf{\bibinfo {volume} {14}},\ \bibinfo {pages} {1005} (\bibinfo {year}
  {2004})%
  \bibAnnoteFile{NoStop}{hoppe}%
\bibitem{holt}%
  \BibitemOpen
  \bibfield{author}{%
  \bibinfo {author} {\bibfnamefont{J.}~\bibnamefont{Holt}}, \bibinfo {author}
  {\bibfnamefont{S.}~\bibnamefont{Singh}}, \bibinfo {author}
  {\bibfnamefont{T.}~\bibnamefont{Drori}}, \bibinfo {author}
  {\bibfnamefont{Y.}~\bibnamefont{Zhang}},\ and\ \bibinfo {author}
  {\bibfnamefont{Z.}~\bibnamefont{Vardeny}},\ }%
  \bibfield{journal}{%
  \bibinfo {journal} {Phys.\ Rev.\ B}\ }%
  \textbf{\bibinfo {volume} {79}},\ \bibinfo {pages} {195210} (\bibinfo {year}
  {2009})%
  \bibAnnoteFile{NoStop}{holt}%
\bibitem{muller}%
  \BibitemOpen
  \bibfield{author}{%
  \bibinfo {author} {\bibfnamefont{J.}~\bibnamefont{M{\"{u}}ller}}, \bibinfo
  {author} {\bibfnamefont{J.}~\bibnamefont{Lupton}}, \bibinfo {author}
  {\bibfnamefont{J.}~\bibnamefont{Feldmann}}, \bibinfo {author}
  {\bibfnamefont{U.}~\bibnamefont{Lemmer}}, \bibinfo {author}
  {\bibfnamefont{M.}~\bibnamefont{Scharber}}, \bibinfo {author}
  {\bibfnamefont{N.}~\bibnamefont{Sariciftci}}, \bibinfo {author}
  {\bibfnamefont{C.}~\bibnamefont{Brabec}},\ and\ \bibinfo {author}
  {\bibfnamefont{U.}~\bibnamefont{Scherf}},\ }%
  \bibfield{journal}{%
  \bibinfo {journal} {Phys.\ Rev.\ B}\ }%
  \textbf{\bibinfo {volume} {72}},\ \bibinfo {pages} {195208} (\bibinfo {year}
  {2005})%
  \bibAnnoteFile{NoStop}{muller}%
\bibitem{sariciftci}%
  \BibitemOpen
  \bibfield{author}{%
  \bibinfo {author} {\bibfnamefont{N.}~\bibnamefont{Sariciftci}}, \bibinfo
  {author} {\bibfnamefont{L.}~\bibnamefont{Smilowitz}}, \bibinfo {author}
  {\bibfnamefont{A.~J.}\ \bibnamefont{Heeger}},\ and\ \bibinfo {author}
  {\bibfnamefont{F.}~\bibnamefont{Wudl}},\ }%
  \bibfield{journal}{%
  \bibinfo {journal} {Science}\ }%
  \textbf{\bibinfo {volume} {258}},\ \bibinfo {pages} {1474} (\bibinfo {year}
  {1992})%
  \bibAnnoteFile{NoStop}{sariciftci}%
\end{thebibliography}%

\end{document}